\documentclass[A4paper,20pt]{article}
\usepackage{multicol}
\usepackage{graphicx}
\usepackage{amssymb,bm,mathrsfs,bbm,amscd}
\usepackage[tbtags]{amsmath}
\usepackage{lastpage}
\usepackage{CJK}
\usepackage{indentfirst}
\date{ }
\begin{CJK*}{GBK}{song}

\title{ Study on the production of the subpicosecond electron bunch }
\author{ Xiongwei Zhu\\
{Institute of High Energy Physics, Chinese Academy of Sciences}\\
{Beijing 100049}}

\begin{document}
\large
\maketitle

\begin{abstract}

   The production of the high brightness femtosecond electron bunch is now one of the hot research topics. This paper describes
one electron linac facility used to produce the subpicosecond electron bunch. We analyze the main structure parameters, and
study the beam dynamics of the facility. Finally, we discuss the application of this facility in Compton scattering.

\end{abstract}



\section{Introdction}
During the past several years, the milestone progress has been made in the field of X-Ray
free electron laser ( XFEL )\cite{e,e0}. Both LCLS at SLAC and SACLA at Spring-8 are successful to produce the hard X-Ray
free electron laser. There arrives the new high peak time for XFEL research. Many new kinds of operation modes
appear, such as echo, self seeding, etc. One of the key techniques for XFEL is the high brightness electron preinjector.
Among the present research cases, the photocathode RF gun is the most successful\cite{d,c}. In case of S-band technique, the BNL type
1.6 cell photocathode RF gun is widely used in many laboratories. This paper does beam dynamics study on the subpicosecond
bunch production facility at Tokyo University ( shown is Figure 1 )\cite{e1,gg,ee}. The linac is made of the photocathode
RF gun, the $2856 MHz$ accelerating structure, the solenoid, and the chicane.

\begin{center}
\includegraphics[width=10cm]{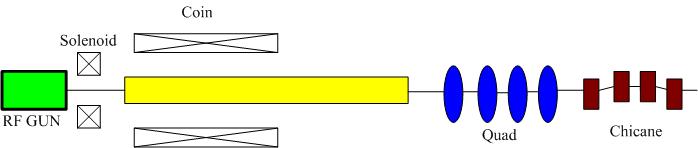}

{ Figure 1.The schematic figure for the linac facility }
\end{center}

\section{Bunch compressing}
Bunch compressing technique\cite{b} is widely used in linear collider and FEL driver linac to get the short high
peak current bunch.  Bunch compressing is usually referred to the magnetic compressing. The compressing
section is made of the RF accelerating part and its related following chicane. In the accelerating part, as the accelerating
phase is not at the peak phase, the related correlated energy spread is produced as the bunch goes through the
accelerating section. The bunch is compressed when the particles with the different energy at the bunch go through
the chicane to form the distance difference. In the early developement, $\alpha$ magnet is often used in the linac
to compress the bunch to drive the low gain FEL oscillator. At present, we commonly use the chicane made of four equal bends to finish the bunch compressing.

\section{Beam dynamics}
The facility is made of the S-band photocathode RF gun, the $2856 MHz$ accelerating structure, and the chicane.
The main parameters of the photocathode RF gun are shown in Table 1.

\begin{center}
{Table 1. The main parameters for the photocathode RF gun}

\begin{tabular}[width=10cm]{@{\extracolsep{\fill}}|c|c|}
\hline
Cathode radius & 1 mm \\
\hline
Quantum efficiency & $>10^{-5}$ at 266 nm \\
\hline
Peak electric field on the cathode surface & 100 MV/m \\
\hline
Energy & 5 MeV \\
\hline
Charge & 100-1000 pC \\
\hline
Eneregy spread & $<0.2\%$ \\
\hline
RF frequency & 2856 MHz \\
\hline
Rep. rate & 10 Hz \\
\hline
Microwave pulse length & 3 $\mu s$ \\
\hline
Microwave peak power& 10 MW \\
\hline
Filling time & 0.55 $\mu s$ \\
\hline
\end{tabular}
\end{center}

The accelerating structure section is made of $61$ cells, and works at $\frac{2 \pi}{3}$ mode.
The main parameters of the accelerating structure are shown in Table 2.

\begin{center}
{Table 2. The main parameters for the accelerating structure}

\begin{tabular}[width=10cm]{@{\extracolsep{\fill}}|c|c|}
\hline
RF frequency &$2856 MHz$   \\
\hline
Mode &$\frac{2 \pi}{3}$  \\
\hline
Field distribution & constant gradient \\
\hline
length & $2.09 m$  \\
\hline
Number of cell & $61$ \\
\hline
Quality factor & $11000$ \\
\hline
Shunt impedance & $57 M\Omega/m$ \\
\hline
Attenuation factor& $0.57 dB$ \\
\hline
Group velocity & $0.02-0.03 c$ \\
\hline
Filling time & 0.55 $\mu s$ \\
\hline
\end{tabular}
\end{center}

The accelerating structure matches the DC electron gun in the oringinal design. Now the photocathode RF gun
take place of the original DC gun, so there is mismatch between the RF gun and the accelerating structure.
The first ten cells are the region of varing phase velocity.  the phase velocity of the first four cells
are $ 0.77c $, the phase velocity of the next three cells are $ 0.94c $, and then the next three cells have the phase
velocity of $ 0.99c $. In addition to the varing phase velocity region, there are $51$ cells with the constant light
velocity. In RF linac, the wave impedance ( the ratio of the transverse electric field to the transverse magnetic
field ) are often defined as

\begin{equation}
Z = 60 \frac{k_{3}}{k} \frac{1 - J_{0}(k_{1}a)}{(k_{1} a) J_{1}(k_{1} a)},
\end{equation}
where $k$ is the wave number in the free space, $k_{3}=\frac{2\pi}{\beta_{w} \lambda}$, $k_{1}^{2} + k_{3}^{2} = k^{2}$, $\beta_{w}$ is the phase velocity. The wave impedance of the varing phase velocity region varies slowly. The structure design
is reasonable. There exists the phase slippage motion in the region of the varing phase velocity due to the mismatch between
the photocathode RF gun and the accelerating structure. The main parameters of the chicane are shown in Table 3.

\begin{center}
{ Table 3. The main parameters for the chicane }

\begin{tabular}[width=12cm]{@{\extracolsep{\fill}}|c|c|}
\hline
Energy &$ 15.46 MeV$   \\
\hline
Energy spread&$ 8.8 \%$  \\
\hline
Compressing ratio& $10$ \\
\hline
$R_{56}$& $-13 mm$  \\
\hline
Total length& $1.04 m$ \\
\hline
Project distance between 1st,2nd bend& $0.194 m$ \\
\hline
Project distance between 2nd,3rd bend & $0.19 m$ \\
\hline
Length of bend& $0.116 m$ \\
\hline
Deflecting angle& $10.4^{\circ}$ \\
\hline
\end{tabular}
\end{center}

Before the chicane, there are four quadrupoles used for the match between the accelerating structure
and the chicane.

We use PARMELA\cite{a} to simulate the linac facility. PARMELA is a multiparticlces PIC code which takes
into account of the space charge effect and is widely used in accelerator community. The version we used
does not include the coherent sychrotron radiation ( CSR ) effect. So the normalized emittance we get in the
simulation is smaller than that in the experiment. We include the fringe field of the bends in the simulation.
There are many choices for the cathode material, such as $Cu$, $Mg$, and $Na_{2}KSb$. When we use $Cu$ or $Mg$,
the main parameters of the laser are shown in Table 4.

\begin{center}
{Table 4. The typical laser parameters}

\begin{tabular}[width=10cm]{@{\extracolsep{\fill}}|c|c|}
\hline
Wavelength & $266 nm$   \\
\hline
Waist & $1 mm$  \\
\hline
Rep. rate& $10 Hz$ \\
\hline
Pulse energy & $20 \mu J$  \\
\hline
Pulse length & $10 ps$ \\
\hline
Rise time & $0.7 ps$ \\
\hline
Longitudinal shape & uniform \\
\hline
\end{tabular}
\end{center}

Figure 2 and Figure 3 give the beam longitudinal distribution before and after the compressing.
It is obvious that the bunch is compressed from $10 ps$ to less than $1ps$. The normalized emittance
dilutes about $0.1 mm mrad$.

\begin{center}
\includegraphics[width=10cm]{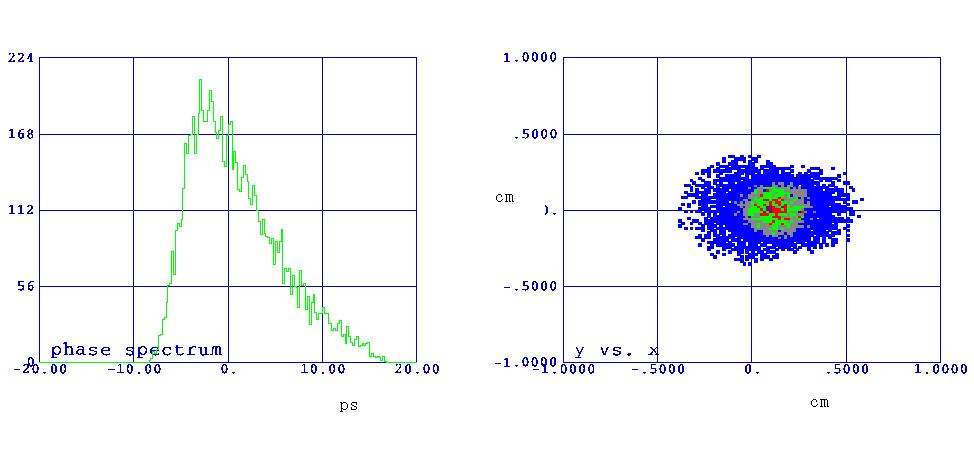}

{Figure 2.The beam longitudinal distribution before compressing }
\end{center}

\begin{center}
\includegraphics[width=10cm]{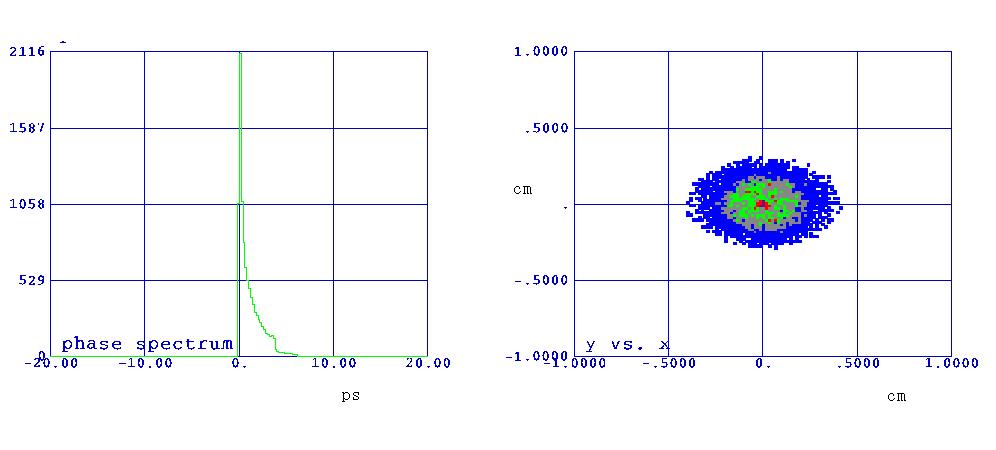}

{Figure 3.The beam longitudinal distribution after compressing }
\end{center}

\section{CSR calculation}
CSR is one of the research topics in the XFEL field. The radiation of the moving charged
particle is a kind of retarded potential. In case of the electron bunch, due to the orbit
arc, the radiation of the bunch tail will have effect on the bunch head. This kind of effect
is called CSR wakefield. When the electron bunch with the energy spread induced by the CSR
impedance goes through the dispersion region, the normalized emittance will dilute. The research
of CSR effect is more complex. The present analytical study is only limited to one dimensional
theory\cite{f}. The high dimensional theory now rely only on the numerical method. Trafic4 is a
better PIC code to simulate the CSR effect. The Trafic4 result shows that $100 pC$ bunch will
have the normalized emittance dilution of $0.3 mm mrad$ going through the chicane. While the
PARMELA result is $0.1 mm mrad$.

\begin{center}
\includegraphics[width=10cm]{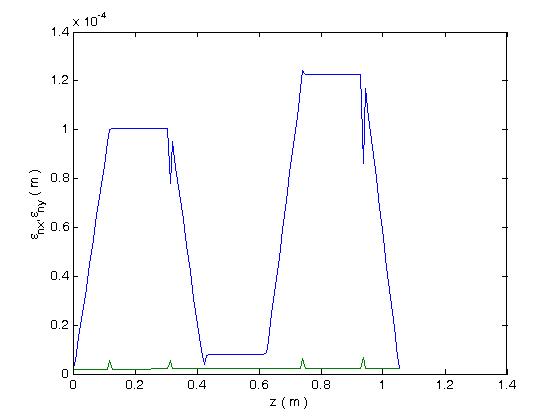}

{Figure 4.The evolution of the normalized emittance ( including CSR effect )}
\end{center}
\section{ Application in Compton scattering}
In comparison with the FEL mechanism, Compton scattering is also a effective method to produce
X-Ray radiation.  Compton scattering is the scattering between the photon and the electron, the energy
of the electron is needless to be high. The coherence and the brightness of the scattering photon is weaker than
that of the free electron laser. We use the following laser parameters ( shown in Table 3) in the simulation.

\begin{center}
{Table 3. The laser parameters used for Compton scattering }

\begin{tabular}[width=10cm]{@{\extracolsep{\fill}}|c|c|}
\hline
Wavelength & $3 \mu m$   \\
\hline
Rayleigh length & $32 \mu m$  \\
\hline
Rep. rate & $10 Hz$ \\
\hline
Peak power & $6.75 \times 10^{14} W/cm^{2}$  \\
\hline
Length & $30 ps$ \\
\hline
\end{tabular}
\end{center}
We use Cain by Yokoya\cite{h} to simulate the Compton scattering photon spetrum. The peak value
is at $1.5 keV$ or so.

\begin{center}
\includegraphics[width=10cm]{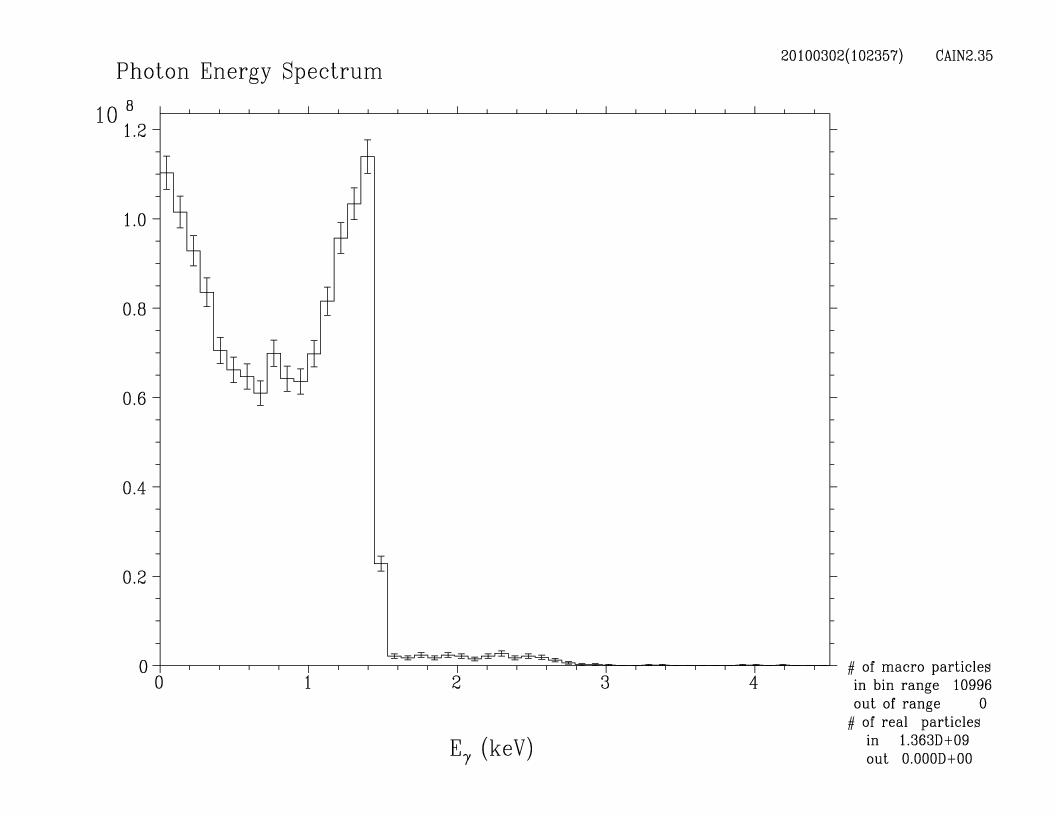}

{Figure 5.The photon energy spectrum }
\end{center}

\section {Discussion}
The low energy linac can also be used to produce X-Ray radiation. In this paper, we simulate
the linac facility used to produce the subpicosecond electron bunch. The produced
electron bunch can be applied both for the X-Ray generation and the inverse free
electron laser acceleration. We simulate the Compton scattering numerically in the
end.

\section*{Acknowledgement}
Thank Dr.T.Kobayashi for the useful discussions.

\end{CJK*}

\begin{thebibliography}{e}
\bibitem{e}{P.Emma, First Lasing and Saturation of LCLS, FEL2009, Liverpool, UK.}
\bibitem{e0}{T.Shintake, First lasing of SACLA, FEL2011, Shanghai, China.}
\bibitem{d}{Xiongwei Zhu, et al, Design Study of a L-band DC Photocathode Gun, Chinese Physics C, V.33, No.4, 311 (2009).}
\bibitem{c}{Xiongwei Zhu, et al, Design Study of a L-band Photocathode RF Injector, Chinese Physics C, Vol.34, No.2, 210 (2010).}
\bibitem{e1}{Xiongwei Zhu, Simulation Study of High Intensity S-band Photoinjector,  APTC Conference, Kansai, Japan (2000).}
\bibitem{gg}{T.Kobayashi,M.Uesaka,Y.Katsumuru,Y.Muroya,T.Watanabe,T.Ueda,K.Yoshii, K.Nakajima, X.Zhu, K.Kando, High Charge S-band
Photocathode RF-Gun and Linac System for Radiation Resaerch, Journal of Nuclear Science and Technology, Vol.39, No.1, 6 (2002).}
\bibitem{ee}{Y.Nakazono, et al, Upgrade of Cartridge-type exchangeable $Na_{2}KSb$ Cathode RF Gun, IPAC10, 4293( 2010 ).}
\bibitem{b}{Xiongwei Zhu, et al, Layout of Bunch Compressor for Beijing XFEL Test Facility, NIM A, Vol.566, 250 (2006).}
\bibitem{a}{L.Young, Parmela Manual (1998).}
\bibitem{f}{E.L.Saldin, et al, Coherent Radiation of an Electron Bunch Moving in an Arc of a Circle, TESLA-FEL-96-24 (1996).}
\bibitem{h}{K.Yokoya, Cain manual ( 2003 ).}
\end{thebibliography}
\end{document}